\documentclass[aps,twocolumn,amsmath,amssymb,showpacs,prl,superscriptaddress,unsortedaddress]{revtex4}
\usepackage{epsf}
\usepackage{graphicx}
\usepackage{sidecap}
\usepackage{amsmath}
\usepackage{float}
 \usepackage{gensymb}
 \usepackage{hyperref}
 \hypersetup{
     colorlinks,
     urlcolor    = blue
}

\usepackage{color}
 
\def \beq {\begin{equation}}
\def \eeq {\end{equation}}
\begin{document}
 
\title{{Observation of topological surface state  in a superconducting material  }}
\author{Gyanendra~Dhakal}\affiliation {Department of Physics, University of Central Florida, Orlando, Florida 32816, USA}
\author{M.~Mofazzel~Hosen}\affiliation {Department of Physics, University of Central Florida, Orlando, Florida 32816, USA}
\author{Ayana Ghosh}\affiliation{Department of Materials Science and Engineering and Institute of Materials Science, University of Connecticut, Storrs, Connecticut 06269, USA}
\affiliation{Theoretical Division, Los Alamos National Laboratory, Los Alamos, New Mexico 87545, USA}
\author{Christopher Lane}\affiliation{Theoretical Division, Los Alamos National Laboratory, Los Alamos, New Mexico 87545, USA}
 \affiliation{Center for Integrated Nanotechnologies, Los Alamos National Laboratory, Los Alamos, New Mexico 87545, USA}

 \author{Karolina G\'ornicka}  \affiliation{Faculty of Applied Physics and Mathematics, Gdansk University of Technology,
Narutowicza 11/12, 80-233 Gdansk, Poland}  
\author{Michal J. Winiarski}  \affiliation{Faculty of Applied Physics and Mathematics, Gdansk University of Technology,
Narutowicza 11/12, 80-233 Gdansk, Poland} 
\author{Klauss~Dimitri}\affiliation {Department of Physics, University of Central Florida, Orlando, Florida 32816, USA}  
\author{Firoza~Kabir}\affiliation {Department of Physics, University of Central Florida, Orlando, Florida 32816, USA}

  \author{Christopher~Sims}\affiliation {Department of Physics, University of Central Florida, Orlando, Florida 32816, USA}
  \author{Sabin Regmi} \affiliation {Department of Physics, University of Central Florida, Orlando, Florida 32816, USA}
\author{William Neff} \affiliation {Department of Physics, University of Central Florida, Orlando, Florida 32816, USA}
\author{Luis Persaud} \affiliation {Department of Physics, University of Central Florida, Orlando, Florida 32816, USA}
\author{Yangyang Liu} \affiliation {Department of Physics, University of Central Florida, Orlando, Florida 32816, USA}
 \author{Dariusz~Kaczorowski}\affiliation {Institute of Low Temperature and Structure Research, Polish Academy of Sciences, 50-950 Wroclaw, Poland}
  \author{Jian-Xin Zhu} \affiliation{Theoretical Division, Los Alamos National Laboratory, Los Alamos, New Mexico 87545, USA}
  \affiliation{Center for Integrated Nanotechnologies, Los Alamos National Laboratory, Los Alamos, New Mexico 87545, USA}

\author{Tomasz Klimczuk}  \affiliation{Faculty of Applied Physics and Mathematics, Gdansk University of Technology,
Narutowicza 11/12, 80-233 Gdansk, Poland} 
 \author{Madhab~Neupane}
\affiliation {Department of Physics, University of Central Florida, Orlando, Florida 32816, USA}
 
\date{6 October, 2019}

\begin{abstract}
\noindent

The discovery of topological insulator phase has ignited massive research interests in novel quantum materials. Topological insulators with superconductivity further invigorate the importance of materials providing the platform to study the interplay between these two unique states. However, the candidates of such materials are rare. Here, we report a systematic angle-resolved photoemission spectroscopy (ARPES) study of a superconducting material CaBi$_2$ [$T_{c}$ = 2 K], corroborated by the first principles calculations. Our study reveals the presence of  Dirac cones   with a topological protection in this system. Systematic  topological analysis based on symmetry indicator shows the presence of  weak topological indices in this material. Furthermore, our transport measurements show the presence of large magnetoresistance in this compound. Our results  indicate that CaBi$_{2}$ could potentially provide a material platform to study the interplay between superconductivity and topology.

\end{abstract} 

\date{\today}
\maketitle
  The advent of topological insulator (TI)  brought a surge of research interests in the topological quantum materials \cite{Hasan, Scjhang, Neupane3}. Three dimensional (3D) topological insulators are non-trivial insulators in
which  bulk behaves as an insulator whereas the
surface state behaves as a metal with spin-polarized
electrons (non-degenerate spin state) dispersing
linearly along the energy-momentum space. The characteristics of the topological insulators
are governed by the  surface state related
to non-trivial topology protected by time-reversal
symmetry \cite{Hasan, Scjhang}. Topological insulators are characterized by Z$_{2}$ invariants which distinguish strong and weak topological insulators from normal insulator. Strong topological insulators have an odd number of Dirac cones in a Brillouin zone, whereas weak topological insulators have an even number of Dirac cones in a Brillouin zone \cite{Hasan, Scjhang}. The discovery of topological insulator invoked the domino effect in disovering the topological semimetals \cite{Hasan, Scjhang, Neupane3, Ashvin} .
 Broadly,  3D topological semimetals are categorized into Dirac semimetal \cite{ cd3as2, Wang}, Weyl semimetal \cite{ TaAs_theory, TaAs_theory_1, Suyang_Science, Hong_Ding} and nodal-line semimetal \cite{ burkov, PbTaSe2, MH3, Schoop, Heng} featured  by  band-crossing, band-degeneracy and symmetry protection \cite{Heng,   SYyang, Wengg}. A 3D Dirac fermion is a massless fermion formed by the crossing of linearly dispersing bulk  conduction and  valence bands at the specific points in  k-space, protected by inversion and time-reversal symmetry. On the other hand, the breaking of  either of the symmetries provides the realization of Weyl semimetallic state. The Weyl nodes  appearing in pairs with two fold degeneracies behave as the source and the sink of the berry phase in the k-space which are connected by the  Fermi arc \cite{Ashvin, Neupane3, SYyang}. Similarly, if the linear band degeneracy persists along the line or ring instead of at discrete points, it is called a topological nodal-line semimetal \cite{burkov, GXu,  CFang}. 

Superconducting state in a topological material promises a possible route to realize the Majorana zero mode (MZM). Similar to other quasiparticles such as Dirac fermion and Weyl fermion, Majorana fermion - a quasiparticle which is the antiparticle of itself - exists in a condensed matter system \cite{MZM1, MZM2}.  Realization of MZM is crucial for the development of  quantum computers as it is anticipated to host  qubit information \cite{MZM2}. Various roadmaps have been proposed in order to realize the Majorana zero mode (MZM) with some success consisting of  strong spin-orbit coupled semiconductor nanowires \cite{Delft}, ferromagnetic  atomic chain \cite{Princeton}, proximity-induced topological superconductor \cite{Hongding1, Hongding2, Vidhya}. However, a robust MZM which can be functional to device fabrication has not yet been realized due to complexities in the fabrications of the heterostructures  \cite{Vidhya}. 
 Therefore, finding a superconducting topological material motivates current research interests.
 \begin{figure*}[ht]
	\centering
	\includegraphics[width=18 cm]{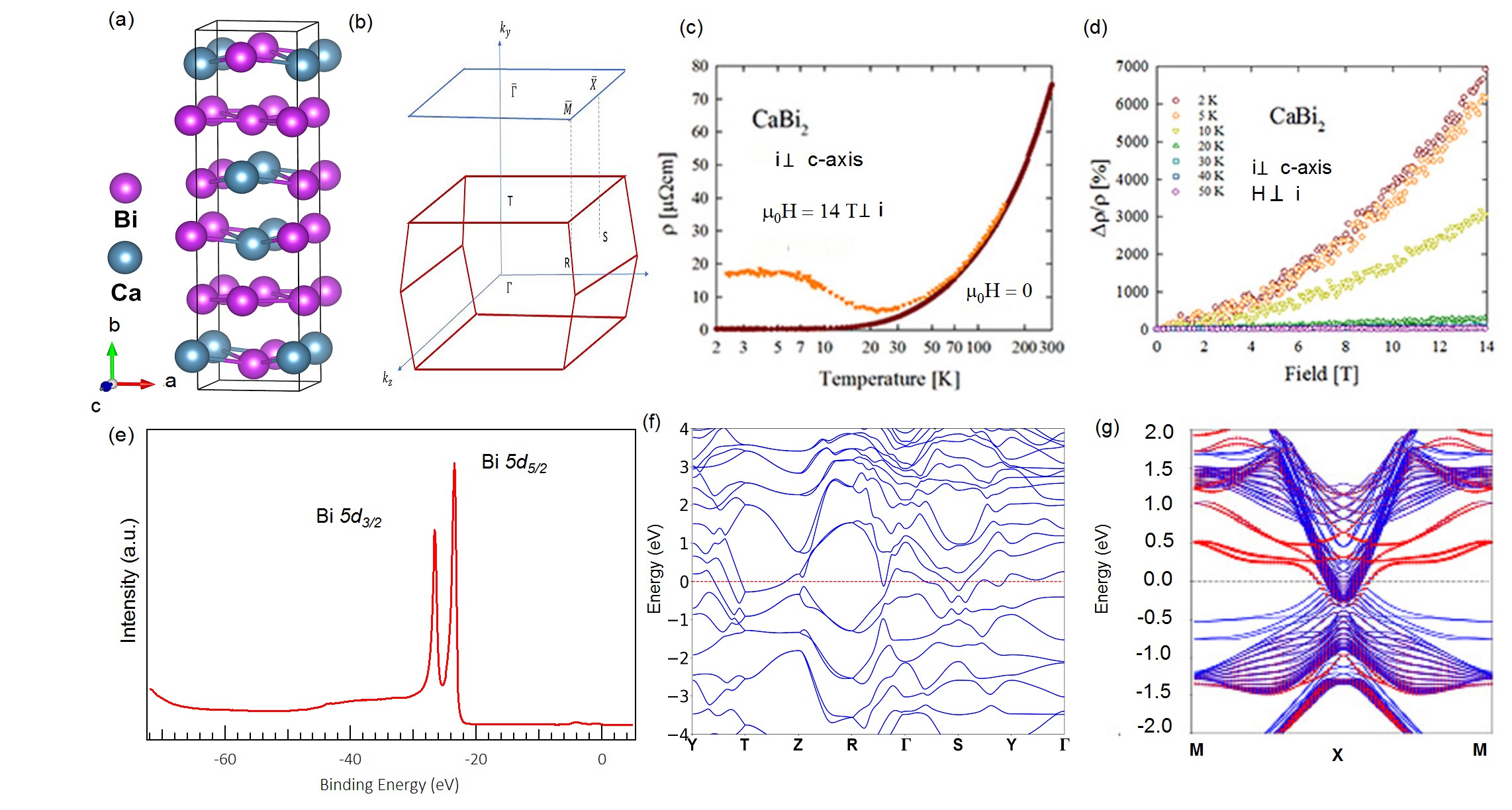}
	\caption{Crystal structure and sample characterization of  CaBi$_{2}$.
(a) Crystal structure of CaBi$_{2}$.  (b) Projection of the 3D Brillouin zone onto the [010] surface, high symmetry points are labeled. (c) Resistivity as a function of temperature at the zero field and under 14 T field. (d) Magnetoresistance as a function of the magnetic field at various temperatures. (e) Inner core spectra   shows the peaks  for Bi 5\textit{d} indicating Bi termination.  (f) First principles calculations of the bulk band of CaBi$_{2}$. (g) Slab calculation of CaBi$_{2}$ along the M-X-M direction.}
\end{figure*}
Lately, tremendous efforts have been invested in order to realize the topological superconductors (TSC) believing it would be a transit to realize MZM. One promising route is to find a material having spin-triplet pairing state, which is commonly known as p-wave superconductor \cite{MZM2}. Cu-doped Bi$_{2}$Se$_{3}$, Sr$_{2}$RuO$_{4}$ are potential candidates of this category \cite{ Liang, Sasaki}. Another sturdy route to realize TSC is the s-wave superconductivity in a spin-polarized state. Some evidences of such systems have been observed, which have  strong Z$_{2}$ index \cite{Suyang_TSC, P}. 
 Recently, large pool of materials with topological insulating state have been discovered, which are defined beyond Fu-Kane criterion using different symmetry settings \cite{Fukane, Hoi, Song, Bernevig}. The interplay between the superconductivity and topology in  these type of novel TI is yet to be established, which provides a new avenue for the research interests.
 
In this communication, we report the electronic band structure of  superconducting CaBi$_{2}$  using angle-resolved photoemission spectroscopy and complemented by first principles calculations. Our study reveals the presence of  Dirac state in CaBi$_{2}$ at the corner point (X) in the Brillouin zone.  Our calculations show the existence of topological insulating state which is embedded within the bulk bands and symmetry indicator analysis also bolsters the existence of weak  topological insulator. Furthermore, this compound demonstrates the large-magnetoresistance indicating topologically non-trivial state.   

\begin{figure} [h!]
	\includegraphics[width=8.7 cm]{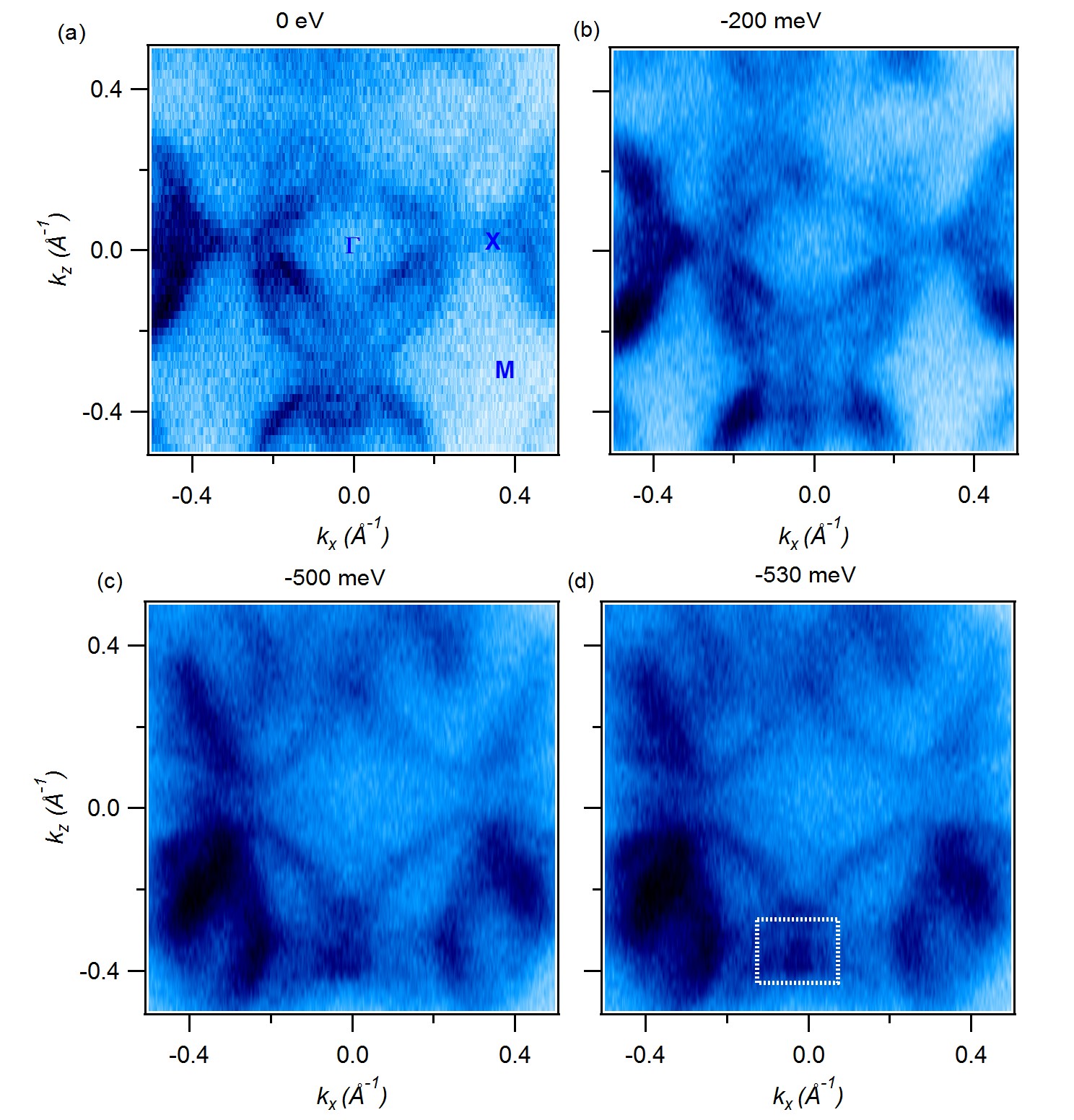}
	\caption{ Measured Fermi surface   and constant energy contours.
(a) Fermi surface map  at a photon energy of  90 eV. High symmetry points are labeled on the plot. (b)-(d) Constant energy contours at different binding energies. Binding energies are noted on the top of each panels. Dotted rectangle serves as a guide to  eyes to visualize the Dirac node.}
\end{figure}
\begin{figure}
	\centering
	\includegraphics[width=9 cm]{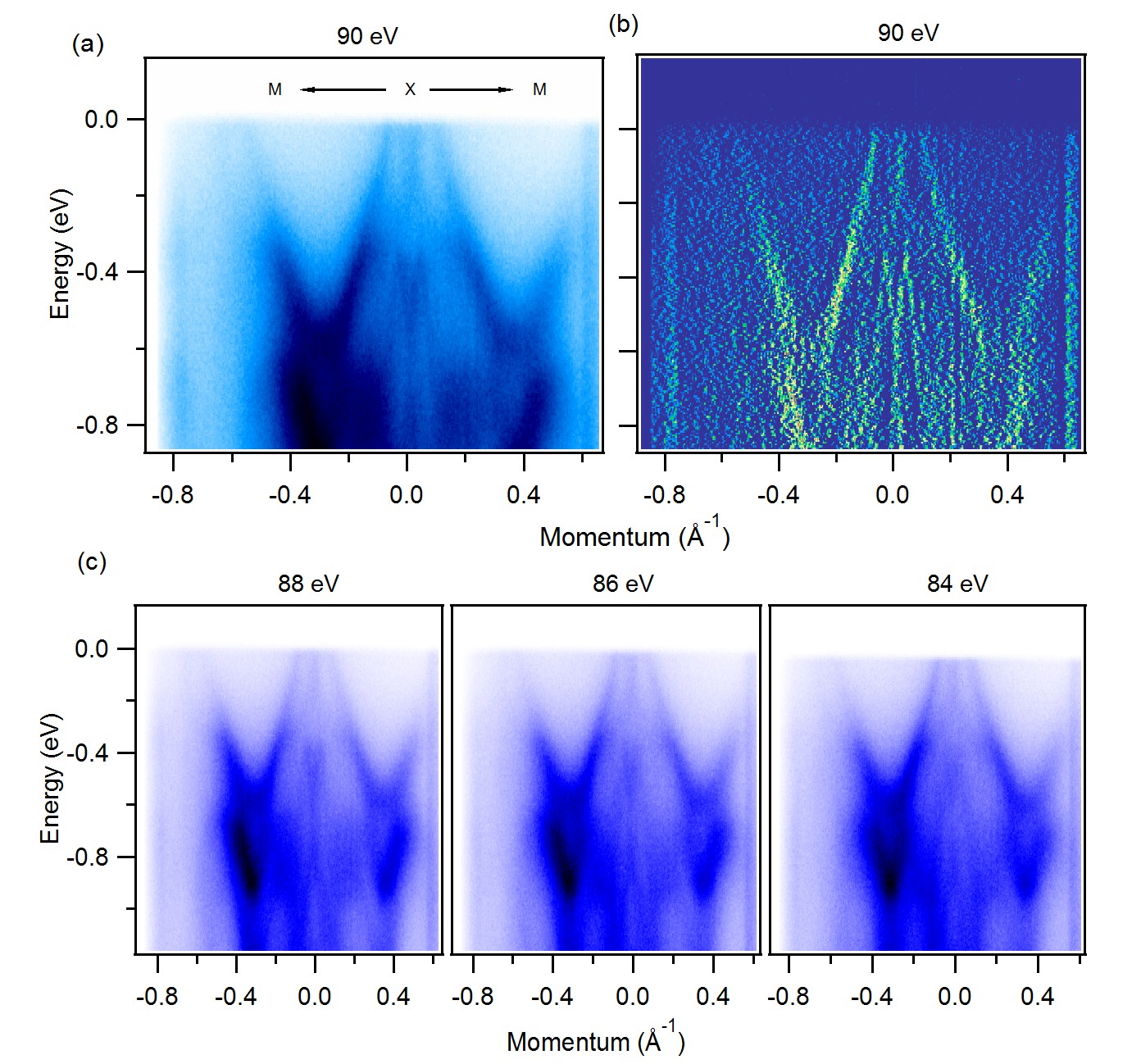}
	\caption{ Photon energy dependent dispersion maps.  (a) Band dispersion along the M-X-M direction measured at a photon energy of 90 eV.  (b) Second derivative plot of dispersion map along the M-X-M direction.  (c) Photon energy dependent band dispersions along the M-X-M direction, photon energies are noted on the top of each figures. All  experimental data were collected at ALS beamline 4.0.3 at a temperature of 20 K.}
\end{figure}
The single crystals of the CaBi$_{2}$ were grown and characterized (X-ray diffraction, energy-dispersive X-ray scepctroscopy, electrical reseistivity, magnetoresistance) as described in Supplemental Material \cite{SM}.  Synchrotron based
ARPES measurements were performed at the Advanced
Light Source (ALS), Berkeley at Beamline 4.0.3
 equipped with a high efficiency R8000 electron
analyzer. The energy resolution was better than
20 meV and the angular resolution was better than 0.2$\degree$.
 Samples were prepared inside a glove box in order to avoid decomposition of the samples. The samples were cleaved in situ and measured between 10 - 80 K in a vacuum better than 10$^{-10}$ torr. The cleavage
plane of single-crystalline CaBi$_{2}$ was the [010] plane.
These crystals were found to be very stable without surface
degradation for the typical measurement period of
20 hours. In order to reveal the nature of the states observed
in CaBi$_{2}$, the ARPES data were compared with
the calculated band dispersion projected onto a 2D Brillouin
zone (BZ). For our first-principles calculations, the WIEN2k package \cite{wien2k} was used (for detailed information see method section in supplemental material \cite{SM}).

 We begin our discussion with the crystal structure of the CaBi$_{2}$. The powdered X-ray diffraction pattern depicts that CaBi$_{2}$ crystallizes in an  orthorhombic lattice, as shown in Fig. 1(a), with non-symmorphic symmetry group \textit{Cmcm}($\#$ 63)\cite{F, MJ}. The crystal structure has quasi-two-dimensional characters, and can be viewed as stacking of a square planar lattice of Bi(1) and a corrugated square lattice of Ca and Bi(2), with the nodes of one lattice positioned above the centers of the squares of the other lattice \cite{MJ}. The lattice parameters are obtained to be a=4.696(1) \AA, b=17.081(2) \AA,  and c= 4.611(1) \AA \quad  which are consistent with  previous reports \cite{F, MJ}. The crystals, on which we performed experiments, are flat, shiny and layered,  with natural cleaving plane of [010]. Therefore, in Fig. 1(b) we present the projection of the 3D Brillouin zone onto [010] surface transforming into a rectangular 2D Brillouin zone.  
 
 As extensively shown in Ref. \cite{MJ}, single-crystalline CaBi$_{2}$ exhibits type-I moderately-coupled superconductivity that emerges below the critical temperature $T_c$ = 2.0 K. Fig. 1(c) presents the temperature dependence of the electrical resistivity measured in the normal state of the very same sample that was investigated in Ref. \cite{MJ}. The experiment was performed with electrical current flowing within the a-c plane of the orthorhombic unit cell. The compound exhibits metallic behavior with the resistivity decreasing in a smooth manner from about 74 m$\Omega$cm at room temperature down to 0.3 m$\Omega$cm at 2 K. Very small magnitude of the residual resistivity as well as large ratio between the values measured at 300 K and 2 K prove that CaBi$_{2}$ is a very good electrical conductor.

Upon applying magnetic field of 14 T, directed perpendicular to the electrical current, the resistivity of CaBi$_{2}$ becomes a non-monotonous function of temperature, namely $\rho$(T) exhibits an upturn below about 30 K and then forms a plateau at the lowest temperatures. Fig. 1(d) presents the transverse magnetoresistance MR = [$\rho$(T,H) - $\rho$(T,0]/$\rho$(T,0) measured as a function of magnetic field at several temperatures up to 50 K. Remarkably, the isotherms taken at 2 K and 5 K, i.e. in the resistivity plateau region, show a nonsaturating behavior with huge values of about 7000-6000 \% in a magnetic field of 14 T. At higher temperatures, the magnitude of MR rapidly decreases; in the same field, it’s only about half that value at T = 10 K, and below 300 \% at T = 20 K. At the highest temperature studied, i.e. 50 K, MR in 14 T drops below 100 \%.

Remarkably, very similar behavior of the transverse magnetotransport has been reported for topologically nontrivial semimetals, like WTe$_{2}$ \cite{Ali}, TaAs$_2$ \cite{Luo} or NbSb$_2$ \cite{Kefeng}, and interpreted either as magnetic-field-driven metal-insulator transition in Dirac systems or an effect of magnetic-field-induced changes in charge carrier densities and mobilites in almost perfectly electron-hole compensated semimetals. It is also worth noting that the magnetoelectrical properties of CaBi$_{2}$ are also very similar to those found for the widely studied nodal-line topological semimetal ZrSiS \cite{Sankar_SR, Singhazrsis} .
  The  core level spectrum displayed in Fig. 1(e) indicates the sharp Bi 5\textit{d} peaks of the cleaved sample.   Our band structure calculations for the bulk CaBi$_2$ (see Fig. 1(f)) show three energy bands crossing the Fermi level along the Y-T direction, suggesting a metallic behavior. However, it is also evident that there is a continuous band gap open between the two energy bands near the Fermi level. 
We used the vasp2trace code  and  
CaBi$_{2}$ is found to be a topological insulator with topological indices Z$_{2w,3}$ = 0, Z$_{4}$ = 0, Z$_{2w,1}$ = 1, Z$_{2w,2}$ = 1 (see supplemental material for details and related references \cite{SM}). For the symmetry group 63, topological indices Z$_4$ and Z$_{2w, i=1,2,3}$   identify the  strong or weak TI. The odd value of former gives the strong TI whereas latter gives the weak TI \cite{Song}. Therefore, on the basis of calculation, we can categorize CaBi$_2$ as a weak TI. We present  slab calculations along the M-X-M direction.  Our theoretical calculations show electron-like bands existing at the X point surrounded by the hole like bands. Furthermore, slab calculations show the bands are enriched with the bulk originations. The linear bands we see in the experiments should be buried within the bulk bands. Our calculations show that the bands-inversion takes place at the S point of the bulk brillouin zone , which projects to the X point on the surface BZ, indicating topological surface state exists at the X point. This system holds an even number of Dirac cones per  Brillouin zone which further confirms the weak TI state in CaBi$_{2}$.

We present ARPES measured electronic structures of CaBi$_{2}$ in Fig. 2.  Fermi surface map and constant energy contours show the band structure evolution with the binding energy.   The  Fermi surface map  reveals the multiple band crossings at the Fermi level which can be seen by multiple Fermi pockets shown in Fig. 2(a).    The spectral intensities of BZs appear unequal because of the matrix element effects, however, the symmetric patterns are well resolved. The Fermi surface exhibits a diamond-shaped Brillouin zone similar to the ZrSiS system \cite{MH3}. The diamond-shaped Fermi surfaces are  intertwined with  other Fermi pockets.  Bands are absent at the zone center ( $\Gamma$), which lies at the center of the diamond-like feature.  The M point makes the center of the diamond shaped feature existing between the Brillouin zones. These two diamond-like features in an extended Brillouin zone make an impression of alternatively changing  tile patterns on the floor.  Presented constant energy contours indicate the presence of the Dirac state showing a Dirac node located at  the binding energy of 530 meV (see Fig. 3 to visualize the Dirac states in different binding energies at the  corner point (X)). A dotted rectangle is drawn in  Fig. 2(d) to depict the Dirac node. Figure 2(b)- (c) show the energy contours at the binding energy of 200 meV and 500 meV respectively showing the evolution of Dirac cone with the binding energy.  At the X point, an electron like band can be seen.  The energy contours clearly show that  hole like bands exist around the $\Gamma$ point, whereas, the M point consists of electron like bands.

In order to understand the nature of the Dirac states in the vicinity of the X point, we present  dispersion maps along the M-X-M direction which reveals the presence of   a linearly dispersive bands state.   
 Linear bands, which cross the Fermi level near to the X point, propagate along the M-X direction. The  potential band crossing between these bands are not accessible to the ARPES measurements due to its location above the Fermi level.  We present a second derivative plot of the intensity map  in Fig. 3(b) which conspicuously shows the presence of  linear bands and Dirac crossings.
   In order to gather an information regarding  the surface nature of the bands along the M-X-M direction, we  provide photon energy dependent dispersion maps in Fig. 3(c). The photon energy dependent  measurements indicate the surface nature of the  Dirac cones.  We present dispersion maps along different  symmetry directions in the Supplemental Material \cite{SM}. 

 Landau-Ginzburg-Wilson framework of spontaneous symmetry
breaking provides the explanation to the exotic phenomena such as superconductivity, magnetism etc \cite{Chiu}. Even though topological systems are beyond this paradigm, the role of the symmetries still plays a guiding role.  Fu-Kane criterion has been well established for distinguishing the topological insulators from trivial insulator.  Z$_{2}$ ($\nu_{0}$; $\nu_{1}$, $\nu_{2}$, $\nu_{3}$) indices define topological indices based on Fu-Kane criterion \cite{Fukane}. Odd value of $\nu_{0}$ indicates strong TI, whereas the odd value of either of $\nu_{i=1,2,3}$ indicates the weak TI.  Recently, symmetry indicators (SI) have expanded beyond Fu-Kane model \cite{Fukane, Song, Bernevig} in order to identify topological materials for different symmetry groups. Fu-Kane criterion segregates TI state from trivial state on the basis of   parity eigenvalues whereas new criterion use symmetry indicators to classify materials. This new criterion has incorporated large number of materials in TI category  providing an opportunity to identify the large number of TI, CaBi$_2$ falls under this category of the materials which cannot be distinguished by Z$_{2}$ indices alone.  The existence of superconductivity in this compound provides us new dimension and perspective.    Due to the strong hybridization among the bands, the surface states are masked by the bulk bands. Therefore, in first-principles calculations the surface states are embedded in  bulk bands.  Being a unique superconducting material from the Ca-Bi family, this stoichiometry is a crucial compound of this family.  Furthermore, this binary superconductor  has high magnetoresistance which is  an indication of topologically non-trivial state in this material. 


In conclusion, our ARPES measurements and first principles calculations reveal the weak topologically protected surface state in a superconducting material CaBi$_{2}$. The surface state is  observed at the X point of the Brillouin zone. The symmetry analysis reveals new class of topological insulating state in CaBi$_{2}$.  Our transport measurements  shows high magnetoresistance possibly arising from non-trivial state.   Our systematic could study potentially provide a new route to study the interplay between superconductivity and topology.

M.N. is supported by the Air Force Office of Scientific
Research under award number FA9550-17-1-0415 and
the National Science Foundation (NSF) CAREER award
DMR-1847962. Work at GUT was supported by the National Science Centre (Poland) under research grant UMO-2017/27/B/ST5/03044. The work at LANL was supported by the U.S. DOE NNSA under Contract No. 89233218CNA000001. J.-X.Z. was supported by the DOE Office of Basic Energy Sciences Program E3B5. C.L. and A.G. was supported  by the Center
for Integrated Nanotechnologies, a DOE BES user facility, in partnership with the LANL Institutional Computing Program for computational resources. M.J.W. was supported by the Foundation for Polish Science (FNP). We thank Jonathan Denlinger for beamline assistance at the ALS, LBNL.

\end{document}